# Long-Lived Spin Coherence States

Yuriy V. Pershin

*Center for Quantum Device Technology,*
*Department of Physics and Department of Electrical and Computer Engineering,*
*Clarkson University, Potsdam, New York 13699-5720, USA*

We study evolution of electron spin coherence having non-homogeneous direction of spin polarization vector in semiconductor heterostructures. It is found that the electron spin relaxation time due to the D'yakonov-Perel' relaxation mechanism essentially depends on the initial spin polarization distribution. This effect has its origin in the coherent spin precession of electrons diffusing in the same direction. We predict a long spin relaxation time of a novel structure: a spin coherence standing wave and discuss its experimental realization.

There is growing interest in the emerging field of spintronics with the aim of controlling and manipulating electron spins in microelectronic devices. Major achievements were attained in the metal spintronics, exploiting the giant magnetoresistance and tunnelling magnetoresistance effects in ferromagnetic-metal-layer and metal/insulator/metal structures [1,2]. A number of metal spintronic devices are already commercialized, for example, magnetic field sensors [3], hard disk recording heads and magnetic random access memory. Significant experimental and theoretical progress in semiconductor spin structures has been reported recently [4-15]. Current research in semiconductor spintronics is mainly focused on spin injection [7] and spin control [8,9], including manipulations of spin coherence time.

Of particular interest in semiconductor spintronics are effects of spin-orbit interaction [16-29]. On one hand, electrically controlled spin-orbit interaction can be used for spin coherence manipulation, as with one of the most prominent device proposals – the spin-field-effect transistor of Datta and Dass [26]. On the other hand, spin-orbit interaction causes electron spin relaxation. The corresponding relaxation mechanism is called D'yakonov-Perel' relaxation mechanism [30,31], arising from bulk asymmetry of a crystal (as with zincblende semiconductors) and/or asymmetry of confining potential. The D'yakonov-Perel' relaxation mechanism is identified as the leading spin relaxation mechanism in many important situations. The asymmetry of confining potential enters into the electron Hamiltonian through Rashba spin-orbit term [32]

$$H_R = \alpha \hbar^{-1} \left( \sigma_x p_y - \sigma_y p_x \right), \qquad (1)$$

where $\alpha$ is the interaction constant, $\vec{\sigma}$ is the Pauli-matrix vector corresponding to the electron spin, and $\vec{p}$ is the momentum of the electron confined in a two-dimensional geometry. From the point of view of the electron spin, the effect of the Rashba spin-orbit coupling can be regarded as an effective magnetic field. In the presence of a magnetic field, the electron spin feels a torque and precesses in the plane perpendicular to the magnetic field direction with angular frequency $\vec{\Omega}$. The quantum mechanical evolution of the electron spin polarization vector $\vec{S} = Tr(\rho \vec{\sigma})$, where $\rho$ is the single-electron density matrix [33], can be described by the equation of motion $d\vec{S}/dt = \vec{\Omega} \times \vec{S}$. Momentum scattering reorients the direction of the precession axis, making the orientation of the effective magnetic field random and trajectory-dependent, thus leading to an average spin relaxation (dephasing). The magnitude of the spin-orbit interaction constant is defined by the geometry of the system, potential gradients and the composition of structure and barriers. The typical value of the effective magnetic field experienced by the electron spin varies from about 10 Gauss in Si-SiGe quantum wells [18] to about 1 T in some III-V compound semiconductor structures [18].

All previous studies of electron spin relaxation in two-dimensional semiconductor heterostructures at zero applied electric field have focused either on properties of spatially homogeneous or spatially inhomogeneous spin polarization but with the same direction of spin polarization vector. In the present paper we report a study of a novel structure – spin coherence standing wave. In this structure the initial direction of spin polarization is a periodic function of coordinate. We show that such a structure is more robust against relaxation then the electron spin polarization having the same direction of spin polarization vector. This phenomenon opens a different approach to spintronic device operation.

*Evolution of a spin polarization strip*. We start our consideration from a simple example, which will help understanding the main idea of our approach. Let us consider evolution of a spin polarization strip. We assume that at the initial moment of time $t=0$ the spin polarization is $\vec{S} = S_0 \hat{z}$ for $|x|<a$ and 0 otherwise ($\hat{z}$ axis is perpendicular to the heterostructure). Initial spin polarization is homogeneous in $y$ direction. The dynamics of electron spin polarization is modeled using a Monte Carlo simulation program described in Ref. [17]. Fig. 1 shows results of our simulations. Evolution of spin polarization components in the central region of the strip is similar to the evolution of the homogeneous spin polarization (see Fig. 1(a)), $S_z$ component decreases with time and $S_x = S_y = 0$. However, behavior of $S_x$ component near the edges of the strip is unusual: it has two pronounced peaks with amplitude comparable to $z$ component of spin polarization. These peaks have the same amplitude but different polarity.

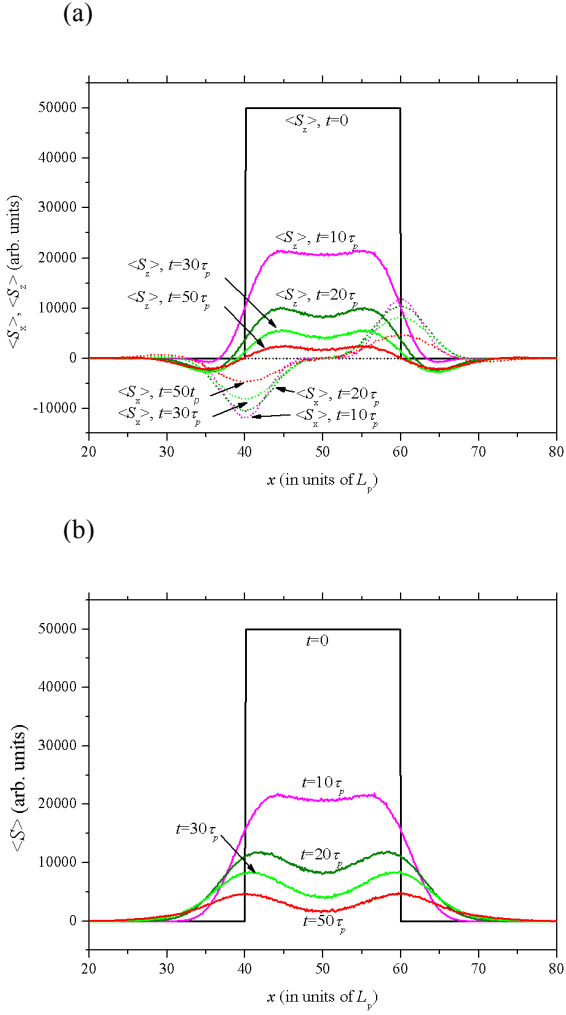

FIG. 1. Evolution of the spin polarization strip: time dependence of spin polarization vector components (a) and amplitude (b).

The amplitude of the spin polarization as a function of coordinate is shown in Fig. 1b. Surprisingly, two peaks of spin polarization are observed at the edge region. This means that relaxation in these regions occurs almost two times slower than in the bulk region. To understand this phenomenon, consider evolution of homogeneous spin polarization. The direction of electron spin precession between two consecutive scattering events is defined by the direction of electron motion. Since the system is homogeneous in $x$ and $y$ directions, the average $S_x$ and $S_y$ spin polarization components of electrons coming to an arbitrary selected space region are zero. When the symmetry of the system is broken, the transfer of spin polarization f from $S_z$ to $S_x$ and $S_y$ may occur. Consider the electrons, for example, near the left edge of initial spin polarization profile. The polarized electrons diffuse out of the area of initial spin polarization, from right to left. Their spins precess coherently, and, since there are no electrons coming to this area from the opposite direction and $S_x$ component becomes uncompensated and conserved, this explains slower spin relaxation in the edge regions. The peaks of $S_x$ in Fig. 1(a) have a different polarity because of the opposite diffusion direction of electrons near the left and right edge. Using the Monte Carlo simulation algorithm we studied spin coherence evolution varying shapes of initial spin polarization profile. It was found that the effect of spin polarization transfer to in-plane components is more prominent with decrease of space dimensions of areas with spin polarization gradients. Similar findings were observed by the authors of Ref. [16], who studied evolution of a pulse of spin polarization.

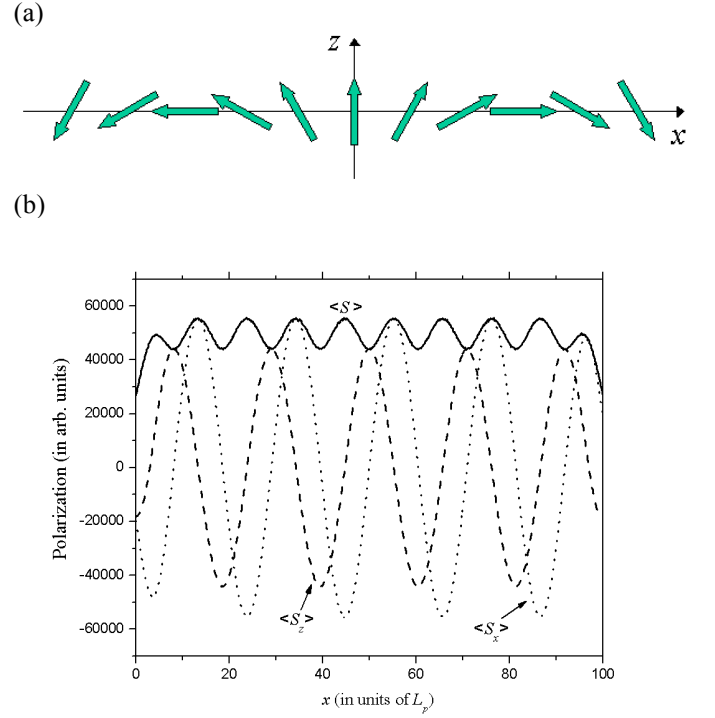

FIG. 2. (a) Schematic of the spin coherence standing wave: direction of spin polarization vector is indicated by the arrows. (b) Total polarization and polarization components of the spin coherence standing wave at $t = 5\tau_p$, with $a = 20.94 L_p$ and $\eta L_p = 0.3$.

*Spin coherence standing wave.* Motivated by observation of longer spin relaxation time near the edge of the spin polarization strip, we study evolution of a spin coherence standing wave, which is schematically shown in Fig. 2(a). Direction of spin polarization in the spin coherence standing wave is a periodic function of $x$ with the components $\left(S_0 \sin(2\pi x/a), 0, S_0 \cos(2\pi x/a)\right)$, where $a$ is the period of the spin coherence standing wave and $S_0$ is its amplitude. Intuitively, the longest spin relaxation time will be, if after passing the distance $a$, the spin precession angle of an electron spin due to the Rashba spin-orbit interaction is equal to $2\pi$. In this case the spin orientation of electrons moving along $x$ direction will coincide with initial direction of spin polarization vector and diffusion along $x$-axis will not lead to relaxation. The electron spin relaxation will be associated only with the electron diffusion in $y$ direction.

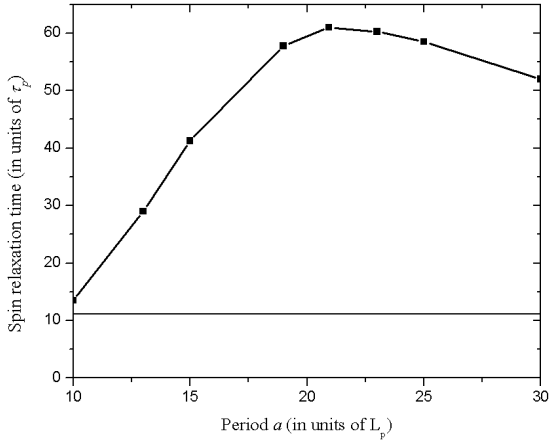

FIG. 3. Dependence of the electron spin relaxation time on the spin coherence standing wave period. The straight line shows the spin relaxation time of homogeneous spin polarization in $z$ direction in the same system.

Fig. 2(b) shows the distribution of the amplitude and components of spin coherence standing wave polarization at some time moment $t > 0$. In our numerical simulations, the spin coherence standing wave was of a finite length, from $x = 0$ to $x = 100 L_p$, which explains the decrease of the spin coherence standing wave amplitude near the edges of this interval. However, we are mostly interested in evolution of spin coherence standing wave in the central region. It is found that in the central region the amplitude of spin coherence standing wave is a periodic function of $x$ with minimums corresponding to maximums of $S_z$ and with maximums corresponding to maximums of $S_x$. We attribute the transition from constant spin polarization amplitude at $t = 0$ to a periodic one at subsequent time moments to the dependence of spin relaxation times on the initial direction of spin polarization vector. It is well known that spin relaxation time of in-plane spin polarization is two times longer then the spin relaxation time of the spin polarization perpendicular to plane [30]. Spin relaxation time of the spin coherence standing wave as a function of its period is depicted in Fig. 3. This dependence has a maximum exactly at $a = 2\pi L_p/\eta$, where $\eta$ is the electron spin precession angle per mean free path. At the maximum, the relaxation time is 6 times as large for the spin coherence standing wave as for the homogeneous spin polarization in $z$ direction. This property of the spin coherence standing wave make it more preferable for technological applications then the homogeneous spin polarization. Another important property of the spin coherence standing wave is its phase. For instance, position of spin coherence standing wave minima could be used to encode the information. An applied electric field in $x$ direction induces sliding of the spin coherence standing wave allowing reading, writing and manipulation of the information.

From the experimental point of view, the spin coherence standing wave can be created in different ways. For example, spin injection from a ferromagnetic metal with rotating magnetization vector to the two-dimensional electron gas can be used. Another possibility is polarization of electrons subjected to a weak magnetic field at different moments of time. Polarization of electron spins is made in the same direction; the applied magnetic field causes precession of electron spins polarized at different moments of time by different angles. Moreover, spin coherence standing wave could be created using superposition of laser beams having distinct polarization and direction.

In conclusion, it was shown that the electron spin relaxation time in two-dimensional systems with inhomogeneous direction of electron spin polarization could be significantly longer as compared to the spin relaxation time in systems with homogeneous spin polarization. A novel structure – spin coherence standing wave characterized by periodicity of direction of spin polarization in one dimension was proposed and studied. Long spin relaxation time of this structure is explained by coherent spin precession of electrons diffusing in the same direction. Two distinctive features of the spin coherence standing wave, namely its long spin relaxation time and its phase, make it attractive for spintronic applications. Possible methods of spin coherence standing wave creation were discussed.

We gratefully acknowledge helpful discussions with Prof. V. Privman and Prof. M. W. Wu. This research was supported by the National Security Agency and Advanced Research and Development Activity under Army Research Office contract DAAD-19-02-1-0035, and by the National Science Foundation, grant DMR-0121146.